\documentclass[english,showpacs,twocolumn,aip,reprint]{revtex4-2}

\renewcommand{\andname}{\ignorespaces} 
\usepackage{mathptmx}
\usepackage[T1]{fontenc}
\usepackage[latin9]{inputenc}
\usepackage{color}
\usepackage{amsmath}
\usepackage{hyperref}
\usepackage{graphics}
\usepackage{natbib}
\usepackage{xcolor} 

\usepackage{graphicx}
\usepackage{epsfig}
\usepackage{bm}
\usepackage{epstopdf}
\usepackage{mathtools}
\usepackage{soul}
\usepackage{subcaption}   
\usepackage{xmpmulti}
\usepackage{lipsum}

\usepackage{babel}
\usepackage{ulem} 

\begin{document}
	
	\title{ On the enhanced Balmer emission of hydrogen in helium Capacitively Coupled Radio Frequency (CCRF) plasma}
	\author{Varsha S}
	\email{varsha.s@ipr.res.in}
	\affiliation{Institute For Plasma Research, Bhat, Gandhinagar,Gujarat, 382428, India}%
	\affiliation{Homi Bhabha National Institute, Training School Complex, Anushaktinagar, Mumbai, 400094, India}

	\author{Prabhakar Srivastav}
	\affiliation{Institute For Plasma Research, Bhat, Gandhinagar,Gujarat, 382428, India}%
	\affiliation{Homi Bhabha National Institute, Training School Complex, Anushaktinagar, Mumbai, 400094, India}
	
	\author{Yukti Goel}
	\affiliation{Institute For Plasma Research, Bhat, Gandhinagar,Gujarat, 382428, India}%

	
	\author{Milaan Patel}
	\affiliation{Physics Department, University of Liverpool, L69 7ZE, United Kingdom}
	\author{Hem Chandra Joshi}
	\affiliation{Institute For Plasma Research, Bhat, Gandhinagar,Gujarat, 382428, India}%
	
	\author{Jinto Thomas}
	\email{jinto@ipr.res.in}
	\affiliation{Institute For Plasma Research, Bhat, Gandhinagar,Gujarat, 382428, India}%
	\affiliation{Homi Bhabha National Institute, Training School Complex, Anushaktinagar, Mumbai, 400094, India}

	\date{\today}
	\begin{abstract}

The present study investigates the observation and enhancement in the intensity of the hydrogen Balmer series emission in a helium CCRF plasma using optical emission spectroscopy (OES).  In addition to the characteristic line emission of helium atoms, the Balmer series of hydrogen and the molecular emission of nitrogen are also observed in the helium discharge. These emissions were primarily attributed to the presence of water vapor in the chamber. In order to confirm the role of helium, the study is also performed using air and argon where no such Balmer series emissions is seen. Experimental evidence suggests that helium metastables transfer energy to trace amount of water content present in the vacuum chamber(present as water vapor). The results point towards the hypothesis that energy exchange between metastable helium and water molecules could be the underlying mechanism. Since the energy of helium metastables exceeds the ionization energy of hydrogen or water vapor molecule, Penning ionization is expected to occur upon their interaction. The hydrogen ions formed as a result, consequently recombine with electrons in the plasma, emitting the Balmer series. 
		Furthermore, the emission intensity of the Balmer series of hydrogen depends on the electron density of the helium plasma. Experiments also show significant deviations in the intensity ratios of the Balmer series from conventional discharges, indicating difference in the underlying population mechanism. A Collisional Radiative (CR) model for measured plasma parameters was used to estimate the metastable population density to understand the mechanism behind the enhancement of the emission intensity. The increase in the metastable densities as well as the radiative recombination cross section appear to be responsible for the observed enhancement. We believe these results will be significant in terms of applications, in addition to providing a fundamental understanding of energy transfer between metastables of helium and water vapor.

	\end{abstract}
	\maketitle
	\section{Introduction}
	
	Capacitively Coupled Radio Frequency (CCRF) plasma has been the focus of considerable research owing to its extensive range of applications and fundamental understanding \cite{raizer2017_ccrf,d2008advanced,Lee_2014,von_Keudell_2017}. Furthermore, CCRF plasma, characterized by moderately high density and temperature in comparison to conventional DC plasma can be considered as a versatile plasma system for the validation of fundamental concepts in collisional radiative (CR) modeling \cite{Bogarts2002}. The plasma parameters estimated through modeling can be readily cross-calibrated using other diagnostics such as electrical probes, enabling precise and comprehensive calibration \cite{Mukherjee2022}. Optical emission spectroscopy (OES), a passive diagnostic method is vital because it does not perturb the plasma. OES using the helium spectral lines is an important diagnostic for the Tokamak edge\cite{Griener2018,Patel2021,Patel2023}. It has been  mentioned that the combination of optical emission spectroscopy (OES) for helium with CR modeling has found significant application as edge diagnostics in fusion-scale machines\cite{Wendler_2022}.
	The physics of collisional energy transfer between atoms and molecules has been an active area of research\cite{Boivin_2005}. Penton et al.\cite{Penton_1968} reported relative cross-sections for the ionization of different molecules using a helium thermal beam. In this work, the authors hypothesized the possibility of Penning ionization to form a collision complex, followed by its dissociation. In a subsequent study, Skoblo et al. \cite{Skoblo_1999} investigated the role of metastable helium atoms and molecules in transferring energy to hydrogen atoms. They examined the time-dependent emission intensities of $H_{\alpha}$ and $H_{\beta}$ and the concentration of metastable atoms in the discharge afterglow to ascertain the dependence of the emission intensity of the hydrogen Balmer series on the concentration of metastable atoms in helium. Another study by Khumaeni et al.\cite{KHUMAENI_2021_OC}, demonstrated that a specific excitation process occurs in microwave-assisted laser plasma induced in helium and argon gases. They observed substantial enhancement of the signal intensity in the helium environment compared to air, which is attributed to the involvement of metastable helium atoms in the excitation process. Lie et al. \cite{C0JA00193G} utilized helium plasma environment to enhance the emission of the Balmer series of hydrogen, thereby enabling effective study of the hydrogen content. A recent study by Philips et al.\cite{ PHILLIPS20087185} showed evidence of the catalytic production of hot atomic hydrogen in a CCRF plasma by mixing hydrogen and helium. They demonstrated the role of molecular hydrogen in the broadening of the $H_\alpha$ emission using a high resolution spectrometer.Qing et al; emphasized the Fulcher-$\alpha$ spectrum in an expanding hydrogen plasma in molecular regime \cite{Qing1996}. This spectra is in the visible range and was used for the estimation of the rotational temperature of neutral gas species. Arora et al\cite{JAAS_Garima} have shown the role of metastable argon on exciting the aluminium neutrals in a laser produced plasma. They showed that the argon metastables are acting as an energy reservoir to exchange energy with neutral aluminium atoms.
	
	\par
	
	Bell et al. \cite{Bell_1968} calculated the interaction energies of helium metastables with different gases and metal atoms and studied the Penning ionization of different species during the interaction. They envisage the formation of an excited complex molecule during the interaction, followed by auto-ionization. In their study, they further pointed to the fact that the auto-ionization probability reaches unity for atoms with less repulsive core, as in the case of hydrogen, compared to other heavier gases such as Nitrogen and Argon.

	The reactions of helium metastable with other species, such as nitrogen, are important to plasma chemistry owing to their ability to produce useful radical species. In a recent study, Myers et al.\cite{Mayers_2024} used optical emission spectroscopy to determine the metastable density of helium in an atmospheric plasma jet of a mixture of nitrogen and helium. They further extended this study to observe that the penning ionization is the dominant ionization channel involving the interaction of metastable helium with N\textsubscript{2} molecules to form the N\textsubscript{2}\textsuperscript{+} radical.Yu et al\cite{Yu_2024_PSST} in a recent work developed a simple model for helium metastable creation, and destruction upon surface impact.

Metastables play a pivotal role in interstellar clouds and planetary atmosphere where the presence of hydrogen and helium are significant \cite{Indriolo2009,Larsson2012}. The interaction with energetic particles gets the helium into metastable stables which later interact with other neutral atoms and molecules to ionize them through the Penning ionization \cite{Queffelec1985,Falcinelli2015}. Penning ionization processes contribute to plasma density variations and alter electromagnetic wave propagation in space plasmas \cite{Falcinelli2015}. Studies have shown enhanced transmission of high-frequency (HF) radio waves following thunderstorms or solar activity, attributed to such increased ionization \cite{Smith1986}.

	It is evident from the literature that helium metastables play an important role in many applications ranging from improving the sensitivity of LIBS for the detection of hydrogen to modulation of plasma chemistry\cite{C0JA00193G,Larsson2012}.Super-sonic molecular beams of helium play a pivotal role in diagnostics and have been identified as an essential tool for edge plasma diagnostics\cite{Patel2021,Patel2023,Wendler_2022}. In this context, the study of the interaction between helium metastables and hydrogen or its isotopes within a plasma environment is of particular importance.  In the present work, we demonstrate that there is an anomalously high intensity of Balmer series emission of hydrogen (hydrogen present in the residual vacuum or adsorbed inside the electrodes or surfaces) in a CCRF plasma of helium. 
	The present study utilizes a double probe measurement technique to elucidate the correlation between the emission intensity and plasma density. 
The manuscript also provides the emission intensity of the Balmer series along the axial direction. The intensity ratio of $H_\alpha$ and $H_\beta$ exhibits a substantial variation from the reported values, depending on the discharge conditions. This has significant importance considering the applicability of this ratio in plasma diagnostics as well as its potential to understand the discharge conditions.
	Additionally, the metastable density of helium is reported from the measurement of OES data using CR modeling. The study provides a systematic investigation of the role of plasma density, discharge conditions, and types of gases and aims to elucidate the enhancement of Balmer series intensity, its dependence on plasma density, and the role of discharge conditions on the intensity ratio of Balmer series.Though there are a few studies on the interaction between metastable helium and hydrogen atoms for different applications, these reported studies did not investigate the details of such interactions particularly on its dependence on plasma parameters and discharge conditions.\par

	\section{Experimental Set-up}
	
\begin{figure}[hbt!]
		\centering
		\includegraphics[width=1.2\linewidth]{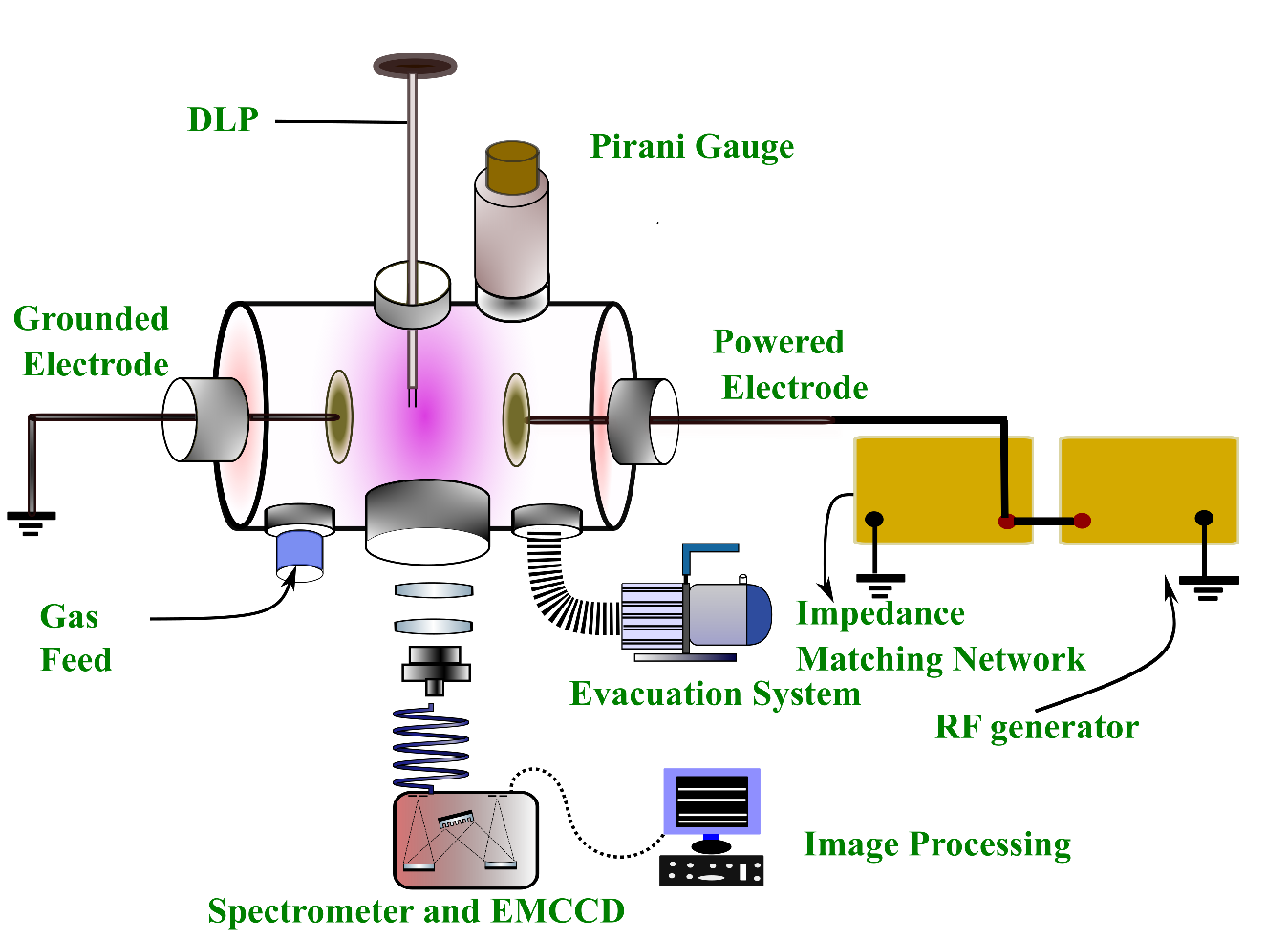}
		\caption{Schematic of experimental setup of CCRF discharge and plasma diagnostics. }
		\label{fig:EXPERIMENTAL_SETUP_MANUSCRIPT}
	\end{figure}

Figure~\ref{fig:EXPERIMENTAL_SETUP_MANUSCRIPT} shows schematic of the CCRF discharge assembly and the plasma diagnostics comprising of a cylindrical glass chamber with a diameter 14 cm and height of 17 cm with multiple ports for diagnostics, housing two brass electrodes each of diameter 4 cm and the associated pumping system to ensure adequate vacuum. A dry pump (Adixen ACP 28), featuring a frictionless pumping module that operates without internal lubricant having pumping speed of 27 $m^3/hr$, provides a rough vacuum of 0.05 mbar inside the chamber. Another turbo molecular pump (Pfeiffer HiPace 300) connected to the system is used to evacuate the chamber to a base vacuum upto $1\times 10^{-4}$ mbar. A precision gas leak valve (Pfeiffer) is used for regulating the gas pressure inside the chamber during the experiment.  A micro Pirani gauge (MKS - model A900-05 ) with built-in calibration for different gases is used for precisely monitoring the pressure of the chamber. The plasma is diagnosed using OES and a Double Langmuir Probe (DLP).

	The plasma is generated using an Advanced Energy CESAR RF Power Supply (model 813977) operating at 13.56 MHz. It couples the output power to the plasma through Advanced Energy Impedance Matching Network (model VM1500AW) in capacitively coupled configuration. The impedance-matching network comprising of inductors and tunable capacitors facilitates efficient power transfer from the RF source to the plasma chamber. The electrodes are mounted on Wilson feed through, allowing  adjustment of the distance between the electrodes and the position with respect to the diagnostics port according to the requirements. 
	
	A 0.25 m spectrograph (Acton SpectraPRO 2356 coupled with an EMCCD Photon MAX 512) is employed for OES .
	An optical assembly comprising of two lenses, with a magnification of 1.2, collects the plasma emissions with a spatial resolution of less than 1 mm when coupled to a fiber bundle array. Care was taken to ensure that stray light is not collected by the imaging system by using proper enclosures and apertures. The orientation of the optical fiber bundle can be changed to collect the OES along the axis of the discharge or perpendicular to it.  The spectrograph is calibrated for the spectral response using a standard white light source and a low pressure Hg-Ar calibration lamp is used for the wavelength calibration. 
	
	A motorized Double Langmuir Probe (DLP) is employed for ascertaining the plasma parameters. The DLP is a floating diagnostic tool that operates by applying a swept voltage between its probe tips to derive electron plasma temperature and density from the resulting current-voltage (I-V) characteristics \cite{Johnson1950}. Although it has certain limitations, the DLP has notable advantages compared to other probe techniques, especially for plasmas that are decaying or exhibit time-dependent potential fluctuations\cite{Sudit1994, Cherrington1982}. Its floating configuration causes minimal disturbance to the plasma, making it a dependable method for measurements under such dynamic conditions.
	The DLP system is mounted on a translation stage enabling to measure the plasma parameters from the desired location accurately. The DLP is made of tungsten wires of 0.8 mm diameter and 6 mm length that are separated by 7 mm. The probe dimensions are selected based on the expected plasma parameters in a typical CCRF plasma. The data has been analyzed using standard symmetric DLP techniques \cite{Brown2017}.
	\section{Results and Discussions}

	\begin{figure}[hbt!]
		\centering
		\begin{subfigure}[b]{0.9\linewidth}
			\centering
			\includegraphics[width=\linewidth]{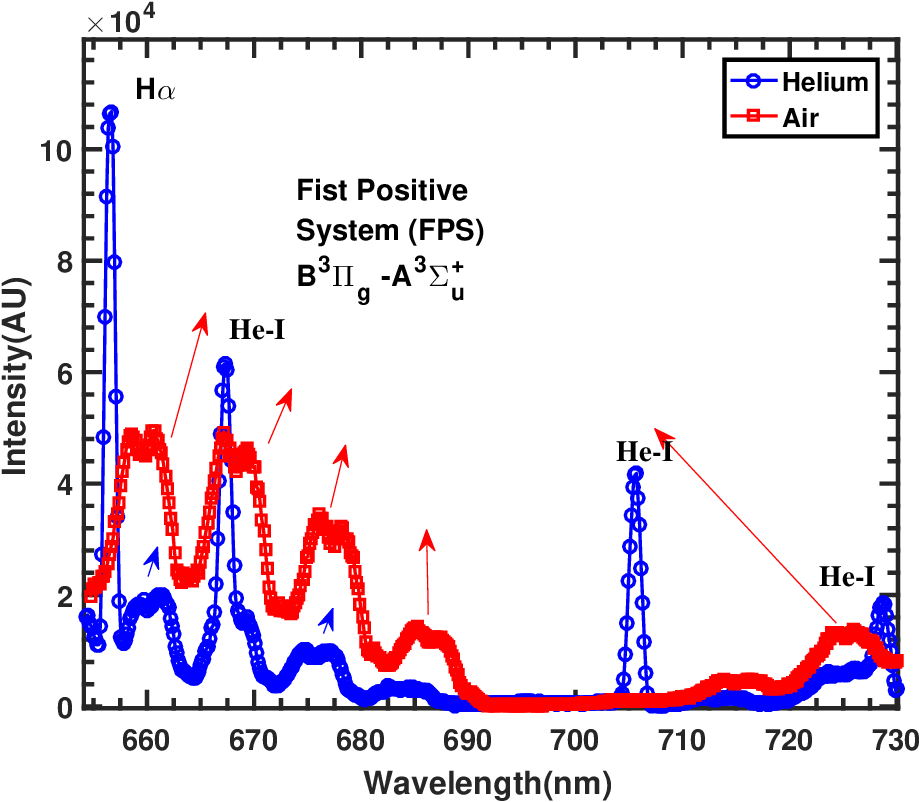}
			\caption{OES of CCRF plasma recorded in the wavelength range of $H_{\alpha}$ and prominent He-I lines.}
			\label{fig:2p587_he_air_comb_5mbar_50w_new}
		\end{subfigure}
		
		\begin{subfigure}[b]{0.9\linewidth}
			\centering
			\includegraphics[width=\linewidth]{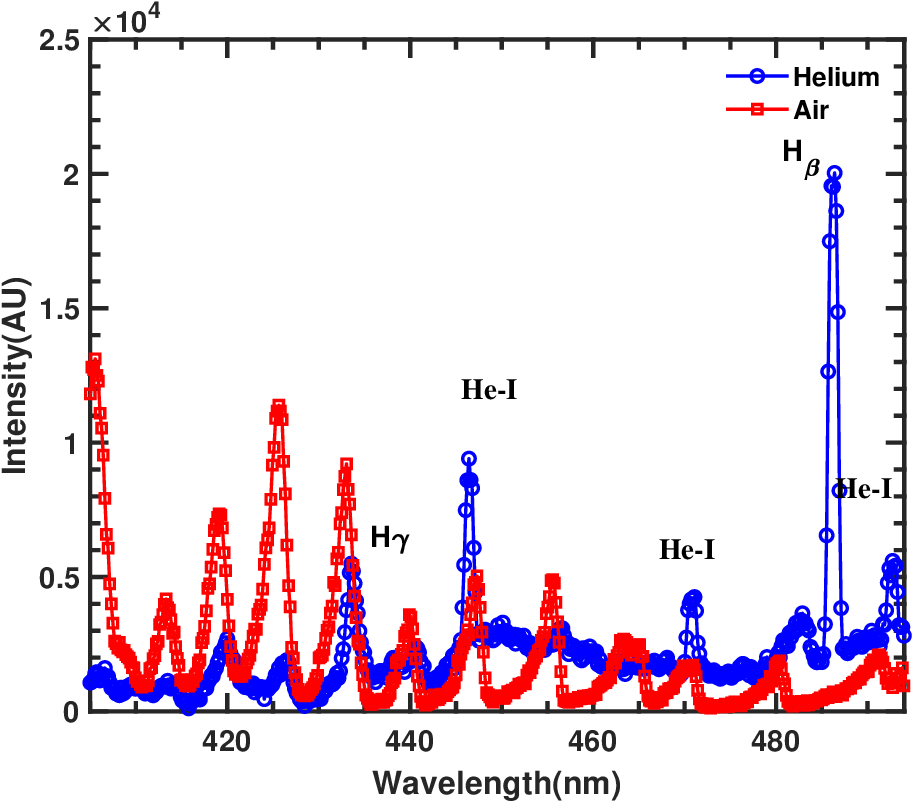}
			\caption{OES of CCRF plasma recorded in the wavelength range of $H_{\beta}$ and $H_{\gamma}$ lines .}
			\label{fig:1p345_he_air_comb_5mbar_50w_new}
		\end{subfigure}
		
		\caption{Optical Emission spectra recorded for CCRF discharge at 50 W RF power at 5 mbar operating pressure for helium and air for two wavelength ranges.}
		\label{fig:combined_emission_spectra}
	\end{figure}
	
	Figures \ref{fig:combined_emission_spectra}
	shows the emission spectra recorded for helium and air plasma at two different wavelength ranges for RF power of 50 W and a pressure of 5 mbar at 16 $\pm $ 1 mm from the RF powered electrode. Each spectrum is an average of 5 frames (2 ms integration time for each frame)to reduce statistical variations. The experiments are repeated to ensure the repeatability of the experiment and found that the intensity variations are with in 6 \% of standard deviation.The OES of  helium plasma shows prominent emissions of He-I, $ H_\alpha $,  $H_\beta$ and molecular emission from $N2$ first positive system (FPS). However, in air discharges the spectra has mostly molecular emissions of $N2$ only.
	The presence of FPS bands of $N_2$ and the Balmer series of hydrogen in helium plasma can be attributed to minor leaks and contamination with water in the experimental chamber. The CCRF chamber made of Glass with o-rings for vacuum sealing has a  limitation in controlling the water vapor content. The base vacuum on the chamber prior to filling helium gas is around $1\times 10^{-4}$ mbar. This is achieved by pumping the vessel using a Turbo Molecular Pump and dry gas purging to reduce the water vapor content. This ensures the partial pressure of water vapor less than 0.1 \% for an experiment with 1 mbar of helium gas. Water vapor in plasma can produce hydrogen atoms or molecules, depending on the plasma parameters. Earlier studies\cite{PHILLIPS20087185,Qing1996,Cook_2025} clearly demonstrated that the presence of molecular hydrogen results in extraordinary broadening of the base of the $H_\alpha$ emission as well as the presence of hydrogen molecular Fulcher bands. However, we did not observe either such a broadening or Fulcher bands, which indicates that dissociation of water vapor results in atomic hydrogen in our case.  As can be seen from the figure ~\ref{fig:1p345_he_air_comb_5mbar_50w_new}, 
	the H$_\gamma$ emission is blended with molecular emission, however H$_\beta$ line is clearly evident. Further, a notable observation is there in the intensity of the Balmer series of hydrogen lines in the helium discharge, which are  stronger than the helium emission lines itself. 
	
	However, OES recorded from the CCRF plasma in air or argon does not show Balmer series of hydrogen, even when normal air with significant water vapor content is used. However, as expected, a substantial enhancement in the intensity of the FPS of nitrogen is evident when atmospheric air used for CCRF discharge.

	The presence of the Balmer series of hydrogen lines in helium discharge, as well as its absence in air and argon discharges using the same power and pressure, is an interesting observation. Therefore, a series of measurements were performed to explore the dependence of RF power and gas pressure on the emission intensities of the Balmer series of hydrogen lines as well as the helium neutral emission lines. 
	
	\begin{figure}[hbt!]
		\centering
		\includegraphics[width=0.9\linewidth]{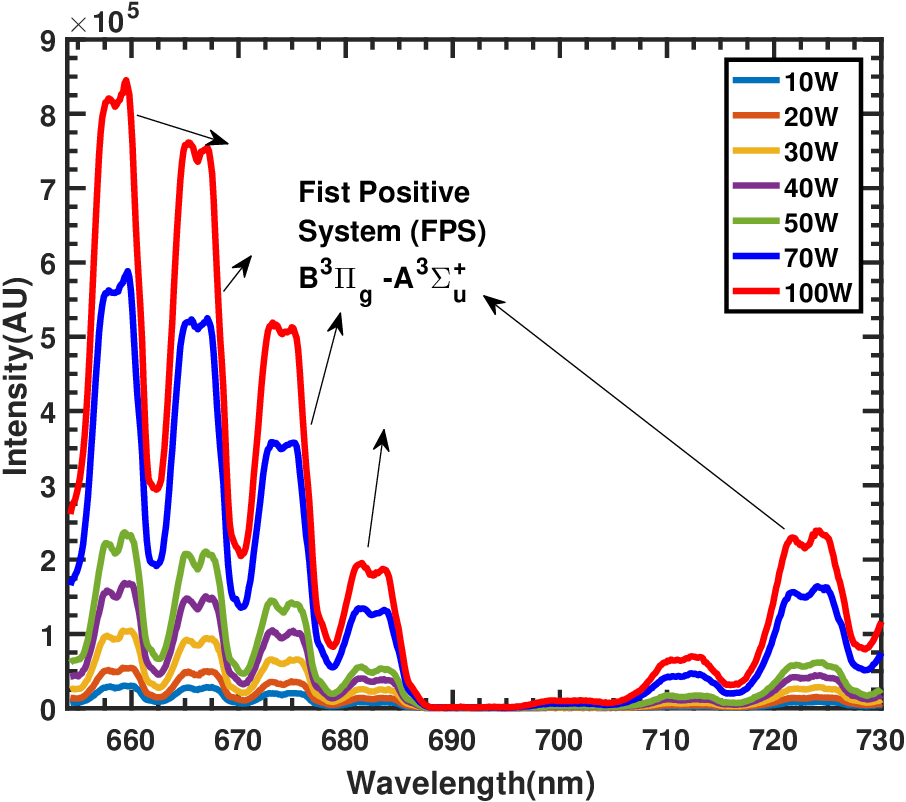}
		\caption{Optical emission recorded for air plasma at 0.5 mbar pressure for various RF powers.  }
		\label{fig:Air_pr5em1_pow_variation_Ha_new}
	\end{figure}
	
	\begin{figure}[hbt!]
		\centering
		\includegraphics[width=0.9\linewidth]{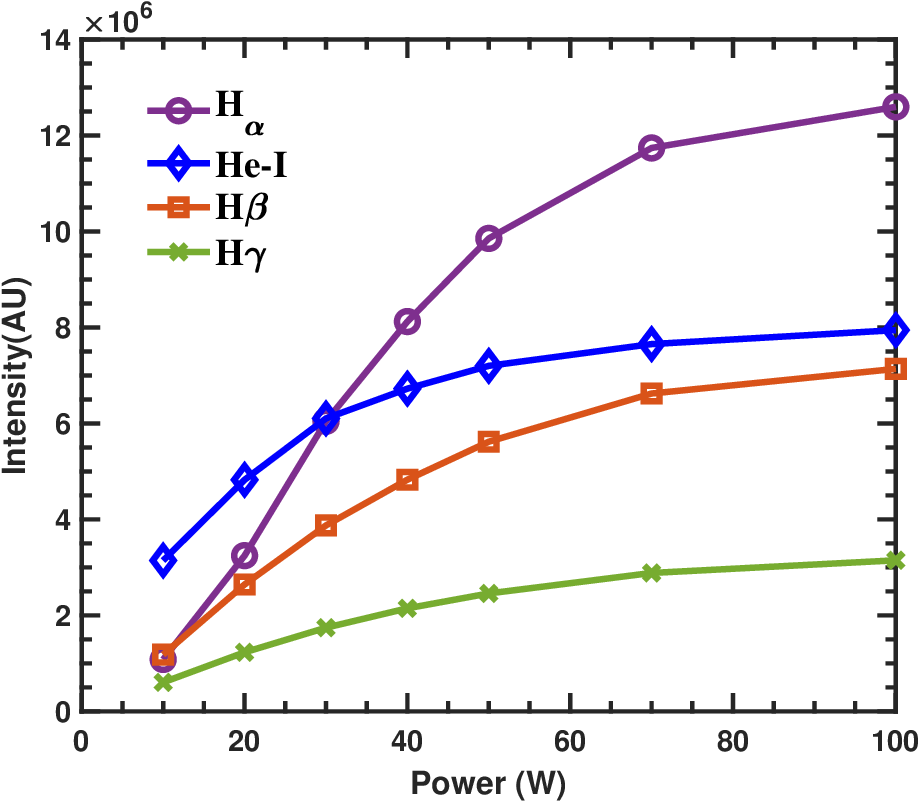}
		\caption{Variation of He-I (706.5),$H_{\alpha}$, $H_{\beta}$ and  $H_{\gamma}$ intensities with power at 0.5 mbar .}
		\label{fig:pow_Ha_He_706_Hb_Hg_combined_16mm_new}
	\end{figure}
	
	Initially, the OES in air discharge is recorded by varying the RF power from 10 W to 100 W at an operating pressure of 0.5 mbar as shown in figure~\ref{fig:Air_pr5em1_pow_variation_Ha_new}. As can be seen from the figure, there is no trace of $H_{\alpha}$ irrespective of the discharge power. However, as the power increases, the emission intensity of FPS of nitrogen increases substantially. Thereafter, similar measurements are performed for helium plasma by varying the RF power. Figure \ref{fig:pow_Ha_He_706_Hb_Hg_combined_16mm_new} shows the intensity variation of the lines of Balmer series of hydrogen ($H_{\alpha}$, $H_{\beta}$ and  $H_{\gamma}$) and helium neutral line (706.5 nm) at a pressure of 0.5 mbar when the RF power is varied. The intensity increases as the power increases and then almost saturates for higher powers. It is worth mentioning here that such a saturation is not seen for the FPS intensity as the power is increased to 100 W. It can be seen from the figure that the intensity of $H_{\alpha}$ was only one-fourth of that of the He I line (706.5 nm) at 10 W of RF power, which increased to nearly twice for RF power of 100 W. Similar behavior in the intensity of other Balmer series of hydrogen lines is also observed, however with a variation in their magnitude of enhancement.

	\begin{figure}[hbt!]
		\centering
		\includegraphics[width=0.9\linewidth]{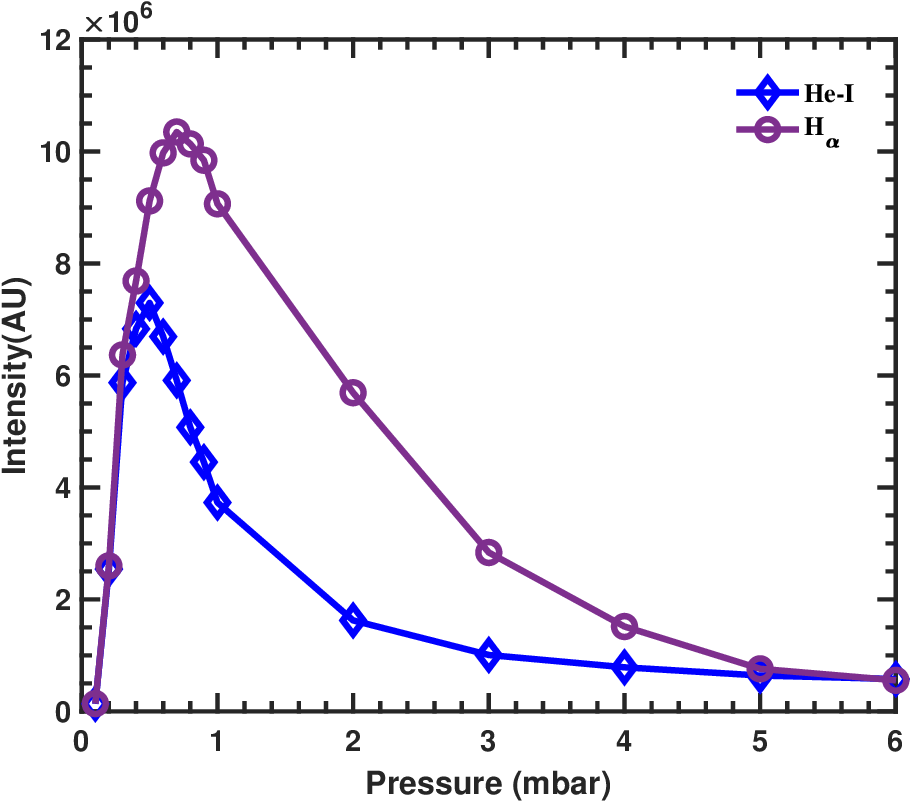}
		\caption{Variation of intensities of $H\alpha$ emission and He-I (706.5 nm) line with helium gas pressure at RF power of 50 W.}
		\label{fig:pr_Ha_new}
	\end{figure}


	
	Similarly, OES is recorded for helium gas by varying the discharge pressure from 0.05 mbar to nearly 6.00 mbar at a given power of 50 W.
	Figure~\ref{fig:pr_Ha_new} shows the variation of emission intensities of $H_{\alpha}$ and He-I lines (706.5) at 16 $\pm $ 1 mm from the RF powered electrode as the pressure increased to 6.0 mbar at 50 W of RF power. The intensities for both the lines increase initially and then starts decreasing. The peaking of intensity for helium emission is around 0.7 mbar, whereas for $H_\alpha$ it is around 1 mbar. Further, the fall in the intensity of helium line is steeper compared to the $H_\alpha$. The results of pressure variation of helium plasma discharge are quite intriguing due to the fact that there is a large enhancement in $H_\alpha$ emission intensity with increase in helium pressure. In fact, the $H_\alpha$ intensity increases $ \sim $75 times when the helium pressure increases from 0.1 mbar to $\sim$ 0.7 mbar much higher than the increase in helium line intensity ($\sim$ 50 times).
	
\begin{figure}[hbt!]
		\centering
		\includegraphics[width=0.9\linewidth]{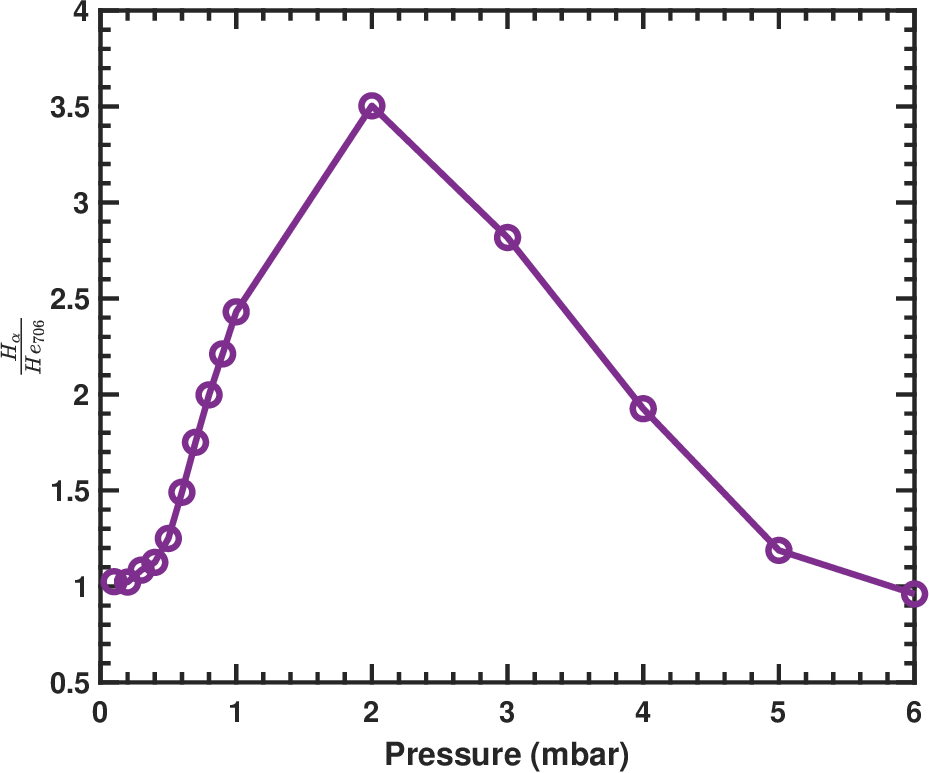}
		\caption{variation of the emission intensity of $H\alpha$ with respect to $He_{706}$ for various  pressure at 50 W of RF power.}
		\label{fig:pr_Ha_He_706_new_16mm_new}
	\end{figure}

	As the emission intensity of a species can depend on various plasma parameters and discharge conditions, normalization of the emission intensity of $H_\alpha$ with respect to the intensity of emission of helium can better demonstrate it. Hence, the ratio of $H_\alpha$ with the helium emission of 706.5 nm (being one of the prominent emission lines of helium) is taken as shown in figure~\ref{fig:pr_Ha_He_706_new_16mm_new}.  The ratio shows an interesting behavior in the relative emission intensities of hydrogen and helium. It is important to note that increase in helium atom density(pressure) increases the emission intensity of $H_\alpha$ more prominently than the intensity of the helium emission line itself. This increase in intensity ratio continues up to a pressure of about 2.0 mbar and then starts decreasing. The decrease in emission intensity of plasma at higher pressures is due to the  dependence on the plasma dynamics and excitation mechanisms. As the discharge pressure increases the mean free path of electrons and its energy reduces. This can decrease the excited state population, subsequently the intensity reduces\cite{LYu_emission_int}. \par
	
	Similar pressure variation study for the discharge in the air as well as in argon gas is also performed, however, we have not observed significant trace of Balmer lines in these cases.

	\par
	
	The experimental observations clearly show that the hydrogen Balmer series emission intensity increases with the pressure of helium gas as well as the applied RF power.  Even though the amount of hydrogen remains the same or in fact decreases depending on its partial pressure upon increasing the helium pressure, the emission intensity of hydrogen Balmer series increases, requires a detailed analysis of plasma parameters. Here, we would like to mention that the helium line intensity ratios with the help of CR modeling are extensively used for the estimation of temperature and density and can be employed to obtain an approximate trends of plasma parameter variations. 
	
	\begin{figure}[hbt!]
		\centering
		\includegraphics[width=1\linewidth]{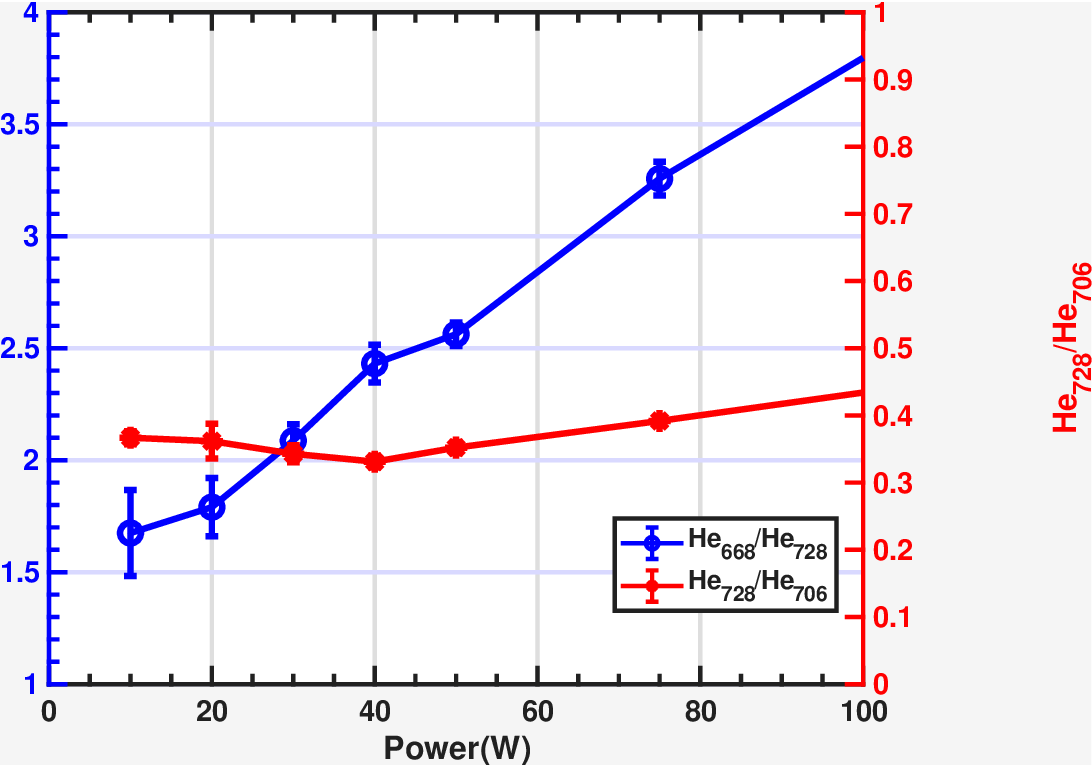}
		\caption{Intensity ratio for plasma temperature(728.1/706.5) and density(668.8/728.1) for He plasma at 5 mbar pressure for various RF power.}
		\label{fig:He_pow_var_temp_dens_ratio_pos2_5em0}
	\end{figure}
	
	\begin{figure}[hbt!]
		\centering
		\includegraphics[width=0.9\linewidth]{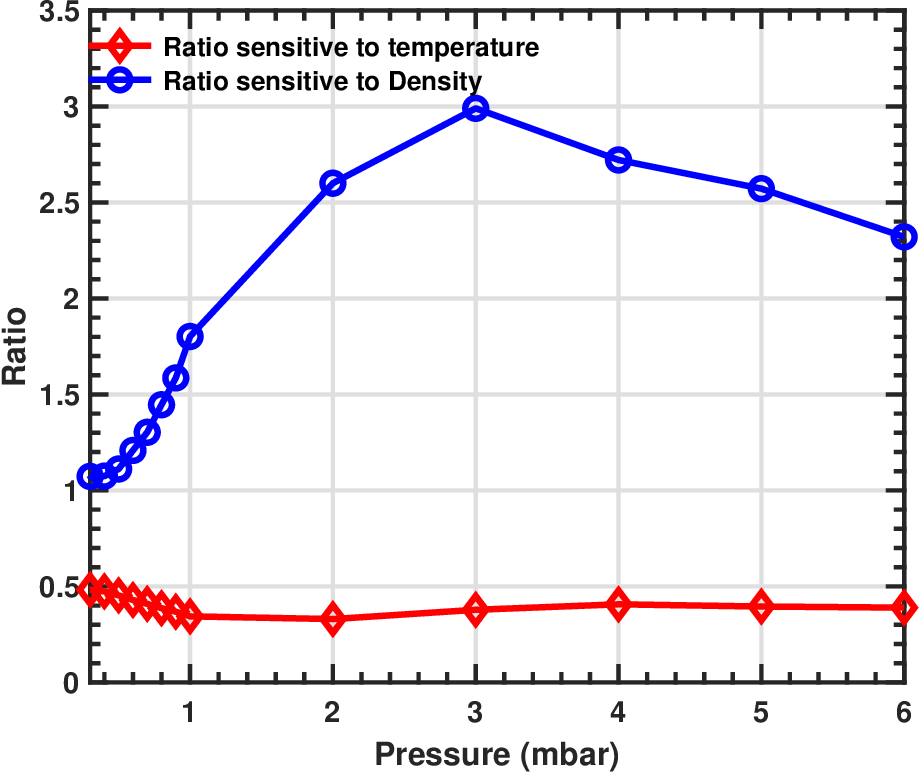}
		\caption{Intensity ratio for plasma temperature(728.1/706.5) and density(668.8/728.1) for He plasma at 50 W Power for various pressures.}
		\label{fig:He_pr_var_temp_dens_ratio_16mm_50W}
	\end{figure}
	
	\begin{figure}[hbt!]
		\centering
		\includegraphics[width=1\linewidth]{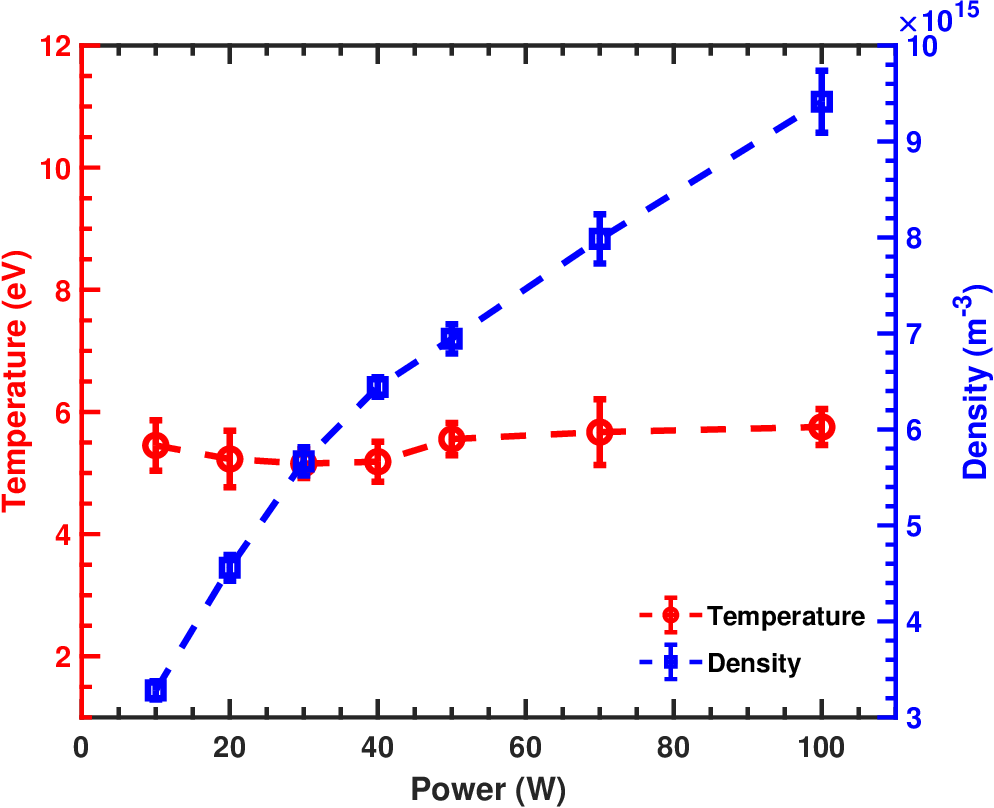}
		\caption{Electron plasma Temperature and density of He plasma estimated using DLP for the RF power at helium gas pressure of 0.5 mbar}
		\label{fig:Power_Temp_Density_He_5em1_dlp}
	\end{figure}

	\begin{figure}[hbt!]
		\centering
		\includegraphics[width=1\linewidth]{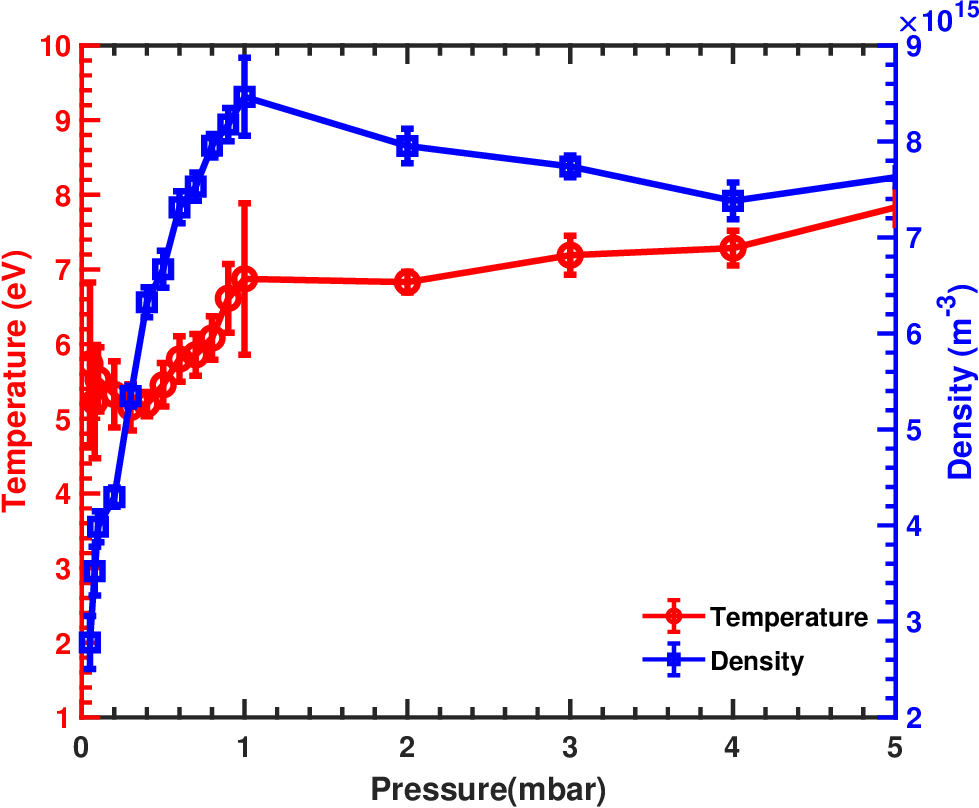}
		\caption{Electron plasma Temperature and density of He plasma estimated using DLP for the variation of helium gas pressures at 50 W RF power}
		\label{fig:Pressure_Temp_Density_He_50W_dlp}
	\end{figure}

	As can be seen from the existing literature, the intensity ratio of 668.8 nm to 728.1 nm is sensitive to the plasma density and the ratio of 728.1 and 706.5 nm is sensitive to the plasma electron temperature\cite{KUBO1999,Goto2014}. Hence for the demonstration purpose, we have plotted these line ratios in figure Figure~\ref{fig:He_pow_var_temp_dens_ratio_pos2_5em0} shows the line intensity ratios of helium lines which are sensitive to electron density and temperature (labeled in the figure) for the RF power variation.  From the figure, it is evident that the ratio of the intensities of the line pair that corresponds to the plasma electron density increases as the power increases. Further, it is interesting to note that the ratio that is sensitive to plasma electron temperature does not change when the RF power increases. Similar line ratios are obtained with pressure variation of helium plasma at 50 W of RF power and are shown in figure~\ref{fig:He_pr_var_temp_dens_ratio_16mm_50W}. Similar to the power variation, the intensity ratio for density shows an increase in the plasma density whereas the ratio sensitive to temperature does not show significant changes as the discharge pressure increases. In fact, the emission intensities (fig~\ref{fig:pow_Ha_He_706_Hb_Hg_combined_16mm_new}) show similar trends with the intensity ratio corresponding to the plasma electron density. Here we would like to mention that although the line intensity ratio is indicative of the plasma parameters, extraction of the accurate values needs extensive CR modeling. In view of this, the temperature and density ranges of present CCRF plasma experiments demands an extensive CR models and have not been pursued at present.

	As discussed, from the line intensity ratios of helium, a possible plasma density dependence on the intensity of emissions of Balmer series of hydrogen is evident. To determine the plasma electron density and temperature, DLP measurement has been done. Figure ~\ref{fig:Power_Temp_Density_He_5em1_dlp} shows the variation of electron temperature and density of helium plasma with RF power at 0.5 mbar of helium pressure from the same location where the OES is recorded. Figure shows that the plasma electron temperature does not vary much with the increase in RF power whereas the plasma electron density is found to increase with the RF power, which is also reflected in the case of helium line intensity ratios. The plasma density increases with an increase in pressure which is well supported by the reported results\cite{HE_2018,TANISLI_2017_153}.  Here it may be noted that the estimated temperature appears higher than what is expected for an RF plasma in the present conditions.
	 As reported by Godyak et al \cite{Godyak_1990} the discharge conditions may change the velocity distribution from maxwellian to Druvestein and can slightly alter the temperature estimation using Langmuir probes. It was also reported that the presence of a small fraction of dust can alter the temperature measured using DLP\cite{Xiong_2023}. 
	Earlier studies on mixture of helium and argon\cite{Naveed2008} demonstrated a significant increase in temperature on increasing the helium concentration, probably a consequence of large non thermal components in helium discharges. A few other studies \cite{Jayraj_RF_plasma,process_12091858} also reported similar temperature range for CCRF discharges. It is important to note that the trend of plasma density and temperature observed are in line with the expected trends reported elsewhere \cite{BORA20121356,HE_2018,TANISLI_2017_153} for the power and pressure variations. As this data is primarily used to observe the trend in variation of plasma parameters, further attempts were not done to refine the parameters obtained from DLP.

It is interesting to note that the trend of $H_\alpha$  emission intensity (figure~\ref{fig:pow_Ha_He_706_Hb_Hg_combined_16mm_new}) and  the plasma electron density with respect to discharge pressure and power is in fact similar.
Hence, the emission intensity of $H_\alpha$ can be correlated to the plasma electron density of  the CCRF plasma. The range of plasma temperature and density for air plasma was also estimated using the DLP and found that it moreover remains the same. Despite the presence of larger water vapor content in the case of air discharges, the Balmer series emission is not observed indicating the intensity of $H_\alpha$ is not simply dependent on the plasma electron density. The absence of Balmer series emission from air and argon plasma with similar plasma electron density and temperature of helium discharges clearly indicate the role of helium atoms on the enhanced Balmer series emission. The helium atoms are known for their metastable states\cite{HOTOP1996191,Porkolab_1975,Endoven1976} and the observed Balmer series emission of hydrogen in the helium discharge may be considered as a consequence of energy transfer from helium metastables to hydrogen.  

	\begin{figure}[hbt!]
		\centering
		\begin{subfigure}[b]{0.9\linewidth}
			\centering
			\includegraphics[width=1\linewidth]{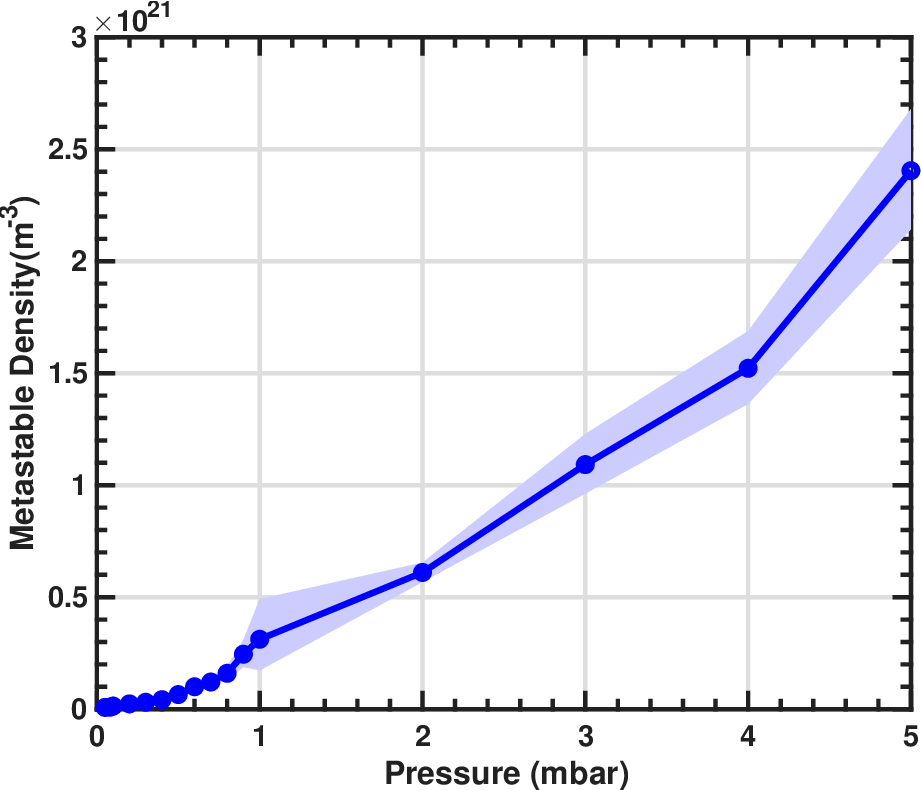}
			\caption{Population density variation of metastable state with increase in pressure estimated using CR modeling for a discharge power of 50 W. The plasma temperature and density measured using double Langmuir probe is used for the estimation}
			\label{fig:metastabledensity_pressure_variation}
		\end{subfigure}
		
		\begin{subfigure}[b]{1\linewidth}
			\centering
			\includegraphics[width=1.05\linewidth]{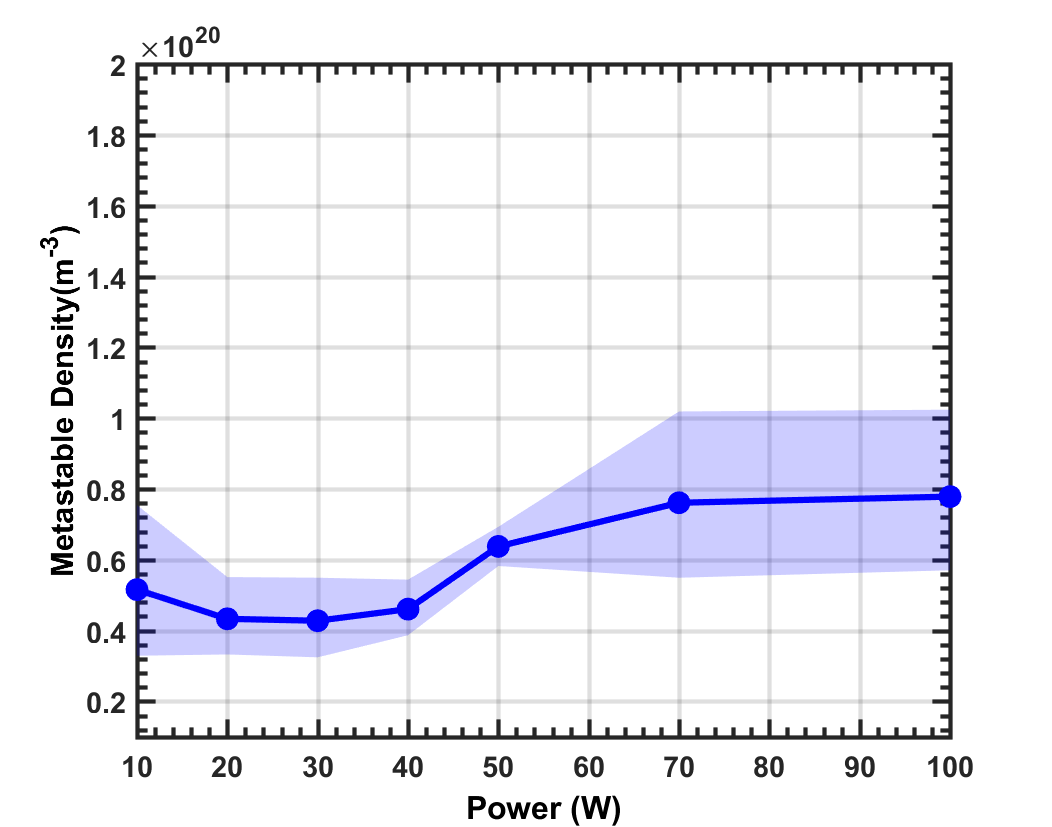}
			\caption{estimated Population density variation of metastables during the discharge power variation at a pressure of 0.5 mbar. }
			\label{fig:metastabledensity_power_variation}
		\end{subfigure}
	\caption{estimated Population density variation of metastables for the power and pressure variations  }	
	\label{fig:Metastable_density}
		\end{figure}

As can be seen from the reported literature, the metastables ${2}^1S$(singlet state) and ${2}^3S$(triplet state) of helium have energies of nearly 20.6 eV and 19.8 eV \cite{HOTOP1996191}respectively.  The population density of the different helium states can be estimated as a function of the population of these metastables states using the open source collisional radiative solver ColRadPy \cite{Colradpy} with measured electron density and temperature (obtained from DLP) as model inputs. To estimate the metastable population densities, we assume the conservation of the total number of helium atoms distributed along all atomic energy levels. Further, this approximation is valid in low-temperature capacitively coupled radio-frequency (CCRF) plasmas, where the degree of ionization is extremely low (of the order of \( \sim 10^{-6} \)). Therefore, the total atomic population at a given pressure \( P \) can be expressed as: $\sum_i N_i = N$ where:
$ N_i$ is the population density of the \( i^\text{th} \) level and $ N $ is the total atomic population at pressure \( P \). Let \( N_2 \) denote the population density of the triplet metastable state and \( N_3 \) that of the singlet metastable state. The population balance can then be rewritten as:
$\sum_{j \neq 2,3} N_j + N_2 + N_3 = N$
The overall population balance can, therefore, be described as the sum of three contributions: the population of all non-metastable states (including the ground state), the population of the triplet metastable state, and the population of the singlet metastable state.
Here, \( N_j \) denotes the population of all other (non-metastable) states, including the ground state.
The population ratios \( N_j/N_2 \) and \( N_j/N_3 \) for each non-metastable level \( j \) can be obtained as function of electron density and temperature. For this , we employed the open source collisional radiative solver ColRadPy. Using these ratios, the summation \( \sum_{j \neq 2,3} N_j \) can be expressed in terms of either \( N_2 \) or \( N_3 \), allowing the system to be reduced to solvable algebraic equations.
 Figure~\ref{fig:Metastable_density} shows the metastable population density variation on varying the discharge power and pressure. Figure clearly shows that as the pressure increases; the metastable density increases significantly whereas with the increase of power the increase in metastable density is not significant.  At a pressure of 0.5 mbar, the dominant populating mechanism for metastables is radiative decay from higher excited states, a process largely independent of electron density and temperature. As a result, increasing the discharge power does not produce a substantial change in metastable density. In contrast, when the power is held constant and the pressure is increased, the electron neutral collision rate increases markedly, thereby enhancing the pathways that populate the metastable states.
As can be seen from the figure that as the discharge pressure increases the metastables density increase significantly.  As the CR model does not include self quenching (as will be discussed latter) of He metastables at higher pressures it shows monotonous increase with pressure.

 As the energy of metastables of helium is much higher than the ionisation energy of hydrogen or water molecule,
	the interaction of metastables of helium with these atoms or molecules results in ionisation, a process known as Penning ionization. It can be mentioned here that the interaction of helium metastables with other atoms and molecules resulting in Penning ionisation is known for a long time\cite{YENCHA1978247,MATTOX2010157}. In the present experimental situation, as already mentioned, the presence of water vapor in the system is the primary source of hydrogen. Different possibilities of formation of hydrogen from water vapor were pointed out in literature\cite{YENCHA1978247,Fantz_2006}.

Some earlier reported studies where mixture of helium and hydrogen are used for experiments, the possible mechanisms of existence of molecular hydrogen or atomic hydrogen are provided. As we are not observing the Fulcher bands of molecular hydrogen as well as deviations in the spectral line shape of $H_\alpha$, we can consider our Balmer series of Hydrogen arising from the interaction of helium metastables with water molecules as given in the following reactions\cite{YENCHA1978247}. 
	\\
		\begin{equation}
	H_2O + He^* \;\longrightarrow\; H_2O^+ + e^- + He
\end{equation}
\begin{equation}
	H_2O^+ \;\longrightarrow\; H^+ + OH^-
	\end{equation}
\begin{equation}
	H^+ + e^- \;\longrightarrow\; H^*
	\end{equation}
\begin{equation}
	H^* \;\longrightarrow\; H + h\nu
	\end{equation}

The dissociation process of $H_2O^+$ can result in excited hydrogen atom through different path ways, which decides the population of the excited states. Depending on the pathways, the emission intensities of the lines of Blamer series may vary.	 
Considering the experimental observations, we believe that Penning ionization should be the reason for the observed Balmer series of hydrogen in the helium discharges. Further, the Penning ionization rate $\Gamma$ depends on several factors, including the density of metastable species $n_A$, target species $n_B$, and the reaction rate$<\sigma v>$ \cite{lieberman2005principles}. The general expression for the ionization rate coefficient can be expressed as
	\[
	\Gamma=n_A*n_B\left\langle \sigma v \right\rangle
	\] 
	
	This clearly shows that as the number of metastables increases, Penning ionization is likely to increase. The possibility of Penning ionization water molecule in a helium plasma is highly expected. The emission of $H_\alpha$ can be attributed to the excited hydrogen atoms, resulting from the recombination. In the observed plasma electron density and temperature ranges of the CCRF plasma, the radiative recombination is expected to be dominated over the three body recombination \cite{Rumsby_1974_Recomb,Mondal2019}. The radiative recombination rate coefficient ($\alpha_R$) can be expresses as follows.

	\[
	\alpha_R=2.7\times 10^{-19}N_e N_i Z^2T_e^{-3/4}
	\] 
	where $Ne,Ni$ are the electron density and ion density in $m^{-3}$ respectively, $Z$ is the charge state and $T_e$ is the electron temperature in eV.
	
\begin{figure}[hbt!]
		\centering
		\begin{subfigure}[b]{0.9\linewidth}
			\centering
			\includegraphics[width=1.0\linewidth]{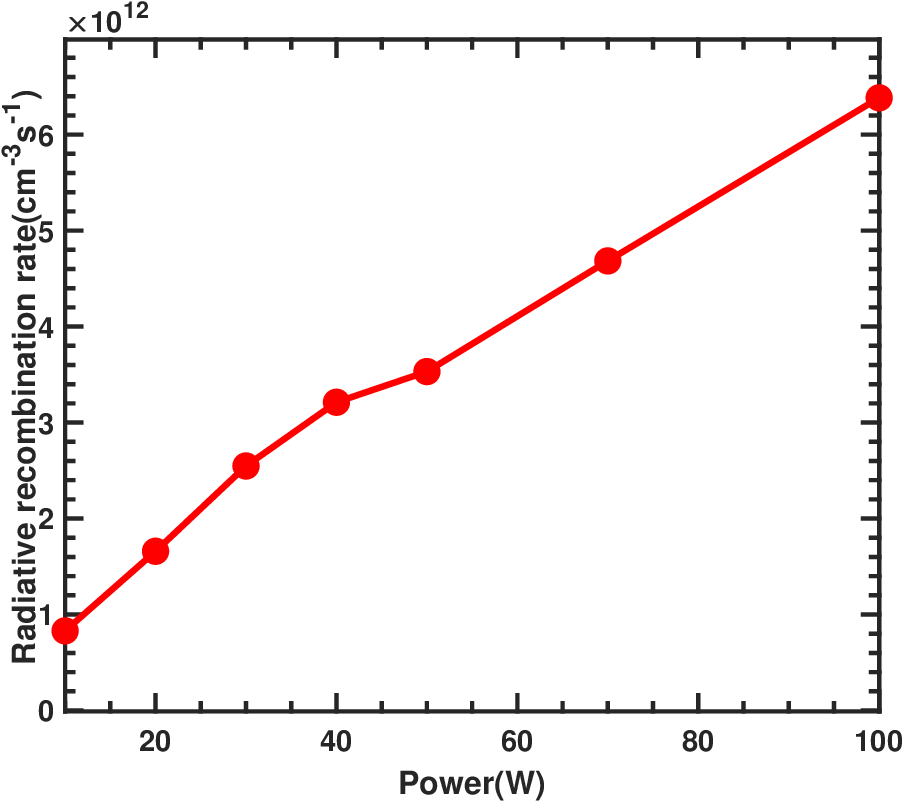}
			\caption{Radiative recombination rates calculated based on the measured temperature and density for different RF powers at a gas pressure of 0.5 mbar.}
			\label{fig:He_5em1_RR_var_power}
		\end{subfigure}
		
		\begin{subfigure}[b]{0.9\linewidth}
			\includegraphics[width=1.0\linewidth]{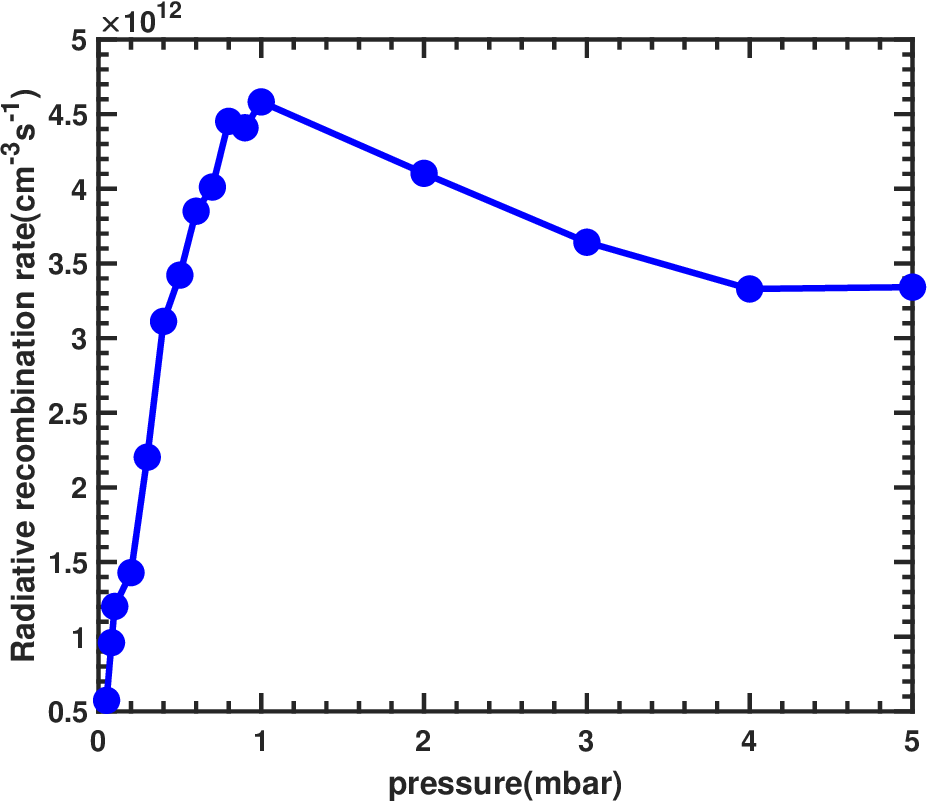}
			\caption{Radiative recombination rates calculated based on the measured temperature and density for different gas pressures of helium plasma at 50 W of RF power}
			\label{fig:He_50W_RR_var_pressure}
		\end{subfigure}
		
		\caption{Radiative recombination rates estimated from the measured temperature and density values for the variation of RF power and gas pressure}
		\label{fig:combined_RR_cross}
	\end{figure}

	The plasma electron density and temperatures measured using the DLP can be used to estimate the radiative recombination rate using the above equation. Figure ~ \ref{fig:combined_RR_cross} illustrates the radiative recombination rates estimated for the measured temperature and density for the pressure and RF power variations. The variation of RF power from 10 W to 100 W shows an enhancement of $\alpha_R$ by nearly 6 times (figure~\ref{fig:He_5em1_RR_var_power}). Furthermore, the enhancement of $H_\alpha$ intensity is found to be approximately 12 times (\ref{fig:pow_Ha_He_706_Hb_Hg_combined_16mm_new})for the similar power variation. The additional factor of 2 in intensity enhancement may be a consequence of more dissociation of H atoms as the plasma density increases or the nominal enhancement in the meatastable population of helium at a  given helium gas pressure as shown in figure~\ref{fig:metastabledensity_power_variation}.

	Figure\ref{fig:He_50W_RR_var_pressure} shows the trend in $\alpha_R$ for the variation of helium pressure at a RF power of 50 W. The trend is similar to the variation of  $H_\alpha$ with helium pressure, thereby pointing to the hypothesis of Penning ionization followed by recombination as the driving force behind the observed presence of the hydrogen Balmer series emission. The enhancement in $\alpha_R$ from 0.05 mbar to 1.0 mbar shows $\approx$ 10 times(figure~\ref{fig:pr_Ha_He_706_new_16mm_new} whereas the emission intensity enhancement for $H_\alpha$ on increasing the helium pressure from 0.05 mbar to 1 mbar is coming to be around 75 times. Here it is to be noted that on increasing the helium pressure the density of metastables increases as shown in figure~\ref{fig:metastabledensity_pressure_variation} and that can result in the additional enhancement in Penning ionization followed by recombination for the increase in $H_\alpha$ signal intensity.
	
 The substantial decrease in the emission intensity of the Balmer series of hydrogen as well as the other line emissions at higher pressures (Fig.~\ref{fig:pr_Ha_new}~\&~\ref{fig:pr_Ha_He_706_new_16mm_new}), similar to the reported results is the consequence of decrease in mean free path of electrons as well as quenching of helium metastables.  The quenching of the helium metastables happens as the background pressure increases. At elevated pressures, the probability of collisions with other molecular species and components of the system, including the chamber walls increases, leading to the quenching of the energy from the helium metastables~\cite{Mayers_2024,Yu2024}.
 Yu et al \cite{Tolmachev2005} reported an enhancement in 
the formation of helium molecule through a three body interaction in a discharge as the pressure increases. It reaches a saturation at around 2 mt tor of discharge pressure of helium an observation similar to we are reporting.  The process involving two ground state helium atom and a metastable helium to form an excited helium molecule and a ground state helium atom. This may result in the decrease in the population of metastables resulting in the decrease in Penning ionization. However, the decrease in the intensity for other emission lines points to the role of plasma conditions and modifications of electron velocity distributions as pressure increases.
 	 The increase in pressure can lead to higher neutral density, which in turn increases the probability of electron-neutral collisions. This increased collision frequency enhances the overall collisions while reducing the effective ionization and excitation cross sections. Consequently, the efficiency of electron-impact ionization and excitation declines, leading to saturation in electron density and a decrease in emission intensity~\cite{Chabert2021,JIANG2022}.

	\begin{figure}[hbt!]
		\centering
		\includegraphics[width=1\linewidth]{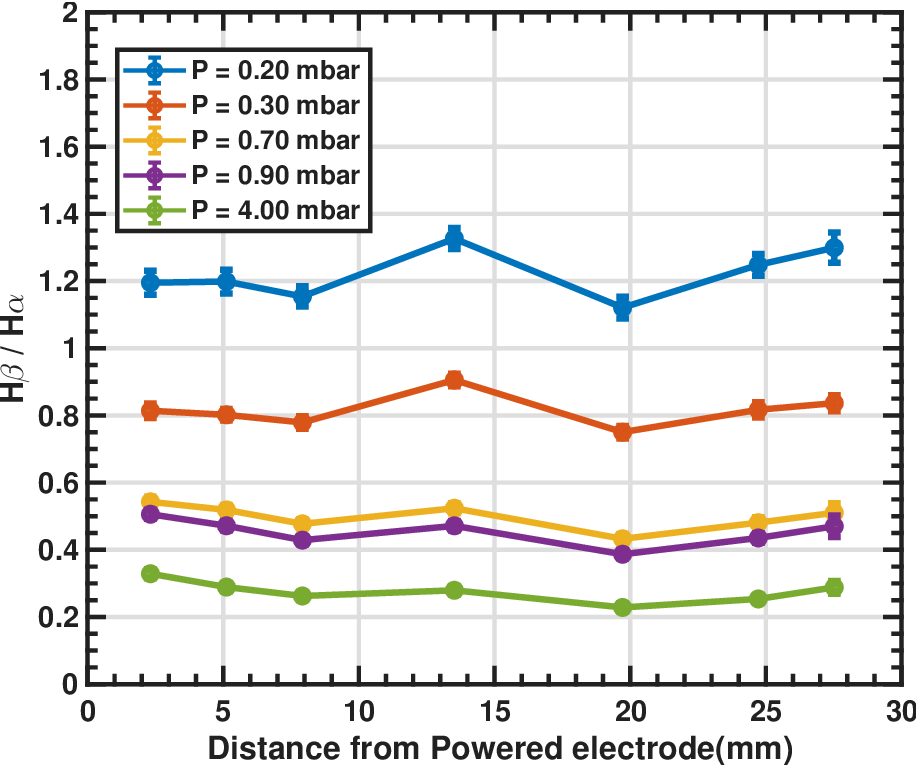}
		\caption{Axial variation of  $H\beta/H\alpha$ ratio at 50 W  of RF power for a range of background pressures.} 
		\label{fig:Hbeta_HALPHA_PRESSURE_new}
	\end{figure}

	\begin{figure}[hbt!]
		\centering
		\hspace{-0.1\linewidth} 
		\includegraphics[width=1.0\linewidth]{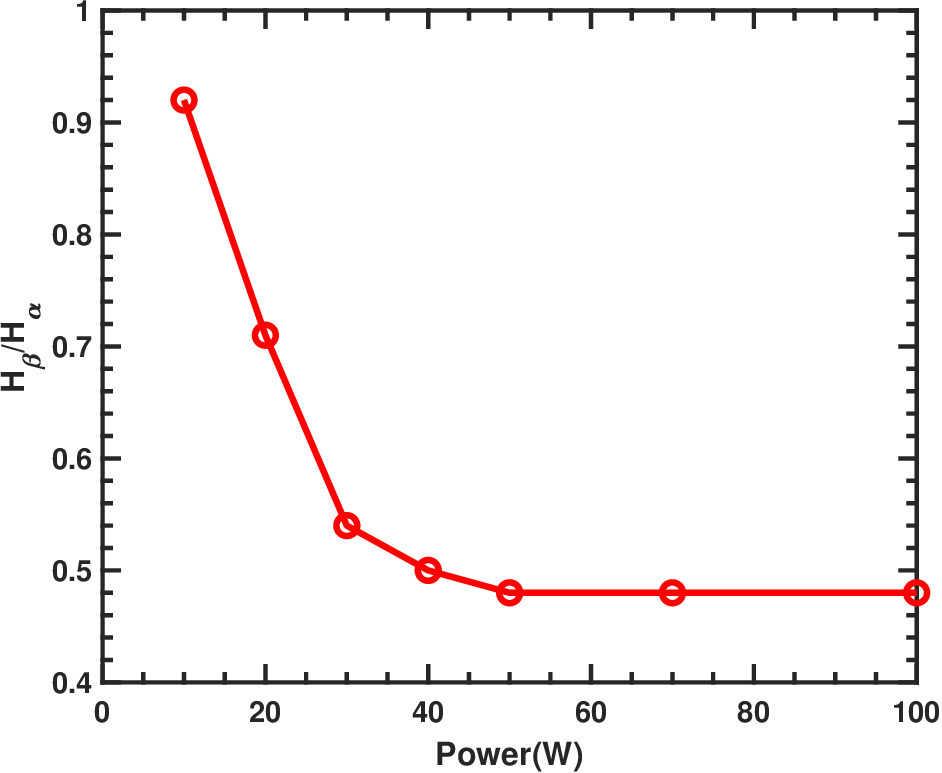}
	
		\caption{Variation of  $H\beta/H\alpha$ ratio with power for a gas pressure of 0.5 mbar }
		\label{fig:Hbeta_HALPHA_ratio_Power_var_5em1}
	\end{figure}
	
One of the most striking observation in this experiment is the observed variations in the ratio of  $H_\alpha$,  H$_\beta$ based on the discharge conditions. For certain discharge conditions, the intensity of  H$_\beta$ is even higher than that of $H_\alpha$. It can be mentioned that the intensity ratio of Balmer series of hydrogen is one of the established diagnostics tool in plasma. Various channels e.g.. recombination, presence of negative hydrogen, dissociative recombination have been discussed in earlier works \cite{Behringer_2000, Fantz_2006,Kakati2017,Verhaegh_2021}.

	Further, in a typical low-pressure discharge plasma of hydrogen, the intensity ratio for $H_\alpha$,  H$_\beta$ and H$\gamma$ is observed to be  1:0.52:0.06 respectively\cite{C0JA00193G,Kakati2017}.However, we observed that the ratio of emission intensities for $H_\alpha$ and H$_\beta$ has a substantial variations depending on the discharge conditions. Figure~ \ref{fig:Hbeta_HALPHA_PRESSURE_new} shows the ratio of  H$_\beta$ to $H_\alpha$ along the axial direction for a few pressures at a discharge power of 50 W. As can be seen from the figure, the ratio varies significantly with helium pressure variation. However,it is important to note that the variation along the axis of the discharge is negligible, even though for the typical RF discharge the plasma density peaks at the center\cite{Godyak_1990}. Hence, it clearly shows the ratio of $H_\alpha$ and H$_\beta$ is not showing any variation with respect to the plasma density. Similar estimate for the ratio at the center of the discharge is also deducted for power variation at 0.5 mbar of pressure and is shown in figure~\ref{fig:Hbeta_HALPHA_ratio_Power_var_5em1}. It can be seen that the ratio varies with RF power significantly. It is interesting as well as unexpected to see that for some instances (a few combinations of powers and pressures) the H$_\beta$ emission intensity is almost equivalent or even higher than that of the $H_\alpha$. The possibility of self absorption for such a skewed ratio is unlikely considering the extremely lower concentration of hydrogen atoms in the system.  Further, this kind of emission behavior is not expected in a system that is in equilibrium. As mentioned earlier, we tentatively suggest that Penning ionization followed by recombination should be responsible for the observed behavior.
	
	Though, the exact mechanism regarding this anomalously higher ratio is not clear, nonetheless this is an important observation for such plasma conditions and needs further exploration. A few studies have reported such deviation in the intensities of H$_\alpha$, H$_\beta$, and H$_\gamma$ lines in the solar chromosphere, as reported by Capparelli et al.\cite{Capparelli2017}. In their study, simultaneous measurements of H$_\alpha$ and H$_\beta$ emissions during the impulsive phase of solar flare activity, where magnetic reconnection events are prominent, revealed that H$_\beta$ intensity was higher than H$_\alpha$. Detailed modeling attributed this behavior to the presence of high heat fluxes generated by non-thermal electron distributions.In the present experimental scenario, at lower discharge pressure similar deviations from the Maxwellian energy distribution for electrons are expected\cite{Godyak_1990}. The presence of such an electron distribution may affect the population of states of hydrogen atom during the radiative recombination. The deviation of emission intensity ratio for the Blamer series of hydrogen is likely to be due to the difference in the mechanism of populations of corresponding upper states. Verhaegh et al\cite{Verhaegh_2021} showed the role of plasma  molecule interaction on altering the ratio of  H$_\alpha$ and H$_\beta$ and its implications for the diagnosis of tokamak divertors using hydrogen atomic line spectroscopy. However, these studies have not considered the Penning ionization route for the formation of excited hydrogen atoms.
	  In the present case the Penning ionization followed by recombination in presence of non maxwellian electron distribution is the population mechanism, whereas in a normal hydrogen discharge it is the impact excitation of hydrogen atoms by electron is the population mechanism. Hence, the observed deviation of $H_\alpha$ $H_\beta$ ratio can differ from the conventional low-pressure discharges. 
	The ratio is higher for lower pressures and power, which points towards the possibility of collisional processes playing a role in the decreasing the population of higher excited states in hydrogen Balmer series. More detailed modeling may be essential to clearly establish the role of recombination and velocity distribution of plasma electrons, which will be pursued further so that it may pave a way for the quantification of the water vapor concentration in a high vacuum system.

	\section {Summary and conclusion}	
	The present study demonstrates the role of helium metastables, plasma density and discharge conditions on enhancing the intensity of Balmer series of hydrogen atoms via Penning ionization followed by recombination in a helium CCRF plasma discharge. The significance of the role of helium metastables in Balmer series emission lines is substantiated by the observations under different discharge conditions using different gases. The presence of hydrogen in helium is expected from  trace amounts of water impurity. However, no such Balmer series emissions are seen in CCRF discharges of air and argon having similar plasma parameters. The CCRF plasma characterized using OES and DLP. The trends in electron plasma temperature and density estimated using DLP are qualitatively verified using the line intensity ratio of helium lines. The CCRF plasma electron density increases with discharge pressure and applied RF power. However, the plasma electron temperature rather remains constant for the power and pressure variations. CR modeling is used to estimate the metastable density of plasma using the experimentally measured parameters.  The intensity of the Balmer series emission is found to correlate with the recombination rates and the metastable concentrations. The radiative recombination rates, estimated with the plasma parameters measured, exhibits a clear trend that is consistent with that of the intensities of Balmer series emission. The intensity ratio of $H_\alpha$ and $H_\beta$ shows a substantial deviation from the conventional emission ratios of hydrogen plasma. The $H_\alpha$ and $H_\beta$ ratio shows that it depends on the discharge conditions rather than the plasma density unlike the intensity of Balmer series emission. Such an intensity ratio has been reported only for a few instances of solar flares related to magnetic reconnection. We believe that the present observations should be of significant importance in understanding the basic energy exchange mechanisms as well as in understanding the role of non-thermal electrons in excitation and re-combinations. Further, rigorous experiments wit precise control of gas species and modeling of the plasma system could provide better understanding of the observed experimental trends.

	\section {Acknowledgment}
	The authors thank Mr Renjith Kumar R, Mr. Vishal Kumar Sharma, Mr Vishnu Chaudhary and Ms B. R. Geethika at Institute for Plasma Research for their technical help.

	
	\bibliographystyle{unsrt}

\end{document}